# Role of Image Acquisition and Patient Phenotype Variations in Automatic Segmentation Model Generalization

Timothy L. Kline, PhD[1,2,*]
Sumana Ramanathan, MS[2]
Harrison C. Gottlich, BS[3]
Panagiotis Korfiatis, PhD[2]
Adriana V. Gregory, MS[2]

[1]Division of Nephrology and Hypertension, Mayo Clinic, Rochester, Minnesota, 55905, USA
[2]Department of Radiology, Mayo Clinic, Rochester, Minnesota, 55905, USA
[3]Alix School of Medicine, Mayo Clinic, Rochester, Minnesota, 55905, USA

**\*Corresponding Author**:
Timothy L. Kline, PhD
Department of Radiology,
Mayo Clinic
200 First St SW,
Rochester, Minnesota, 55905
USA
Email: kline.timothy@mayo.edu

**Competing Interests**
None of the authors have financial interests to disclose and the authors declare no conflicts of interest.

**Funding information**
This work was supported by the NIDDK [grant numbers R03DK125632, K01DK110136].

ABSTRACT

**Purpose:** This study evaluated the out-of-domain performance and generalization capabilities of automated medical image segmentation models, with a particular focus on adaptation to new image acquisitions and disease type.

**Materials:** Datasets from both non-contrast and contrast-enhanced abdominal CT scans of healthy patients and those with polycystic kidney disease (PKD) were used. A total of 400 images (100 non-contrast controls, 100 contrast controls, 100 non-contrast PKD, 100 contrast PKD) were utilized for training/validation of models to segment kidneys, livers, and spleens, and the final models were then tested on 100 non-contrast CT images of patients affected by PKD. Performance was evaluated using Dice, Jaccard, TPR, and Precision.

**Results:** Models trained on a diverse range of data showed no worse performance than models trained exclusively on in-domain data when tested on in-domain data. For instance, the Dice similarity of the model trained on 25% from each dataset was found to be non-inferior to the model trained purely on in-domain data.

**Conclusions:** The results indicate that broader training examples significantly enhances model generalization and out-of-domain performance, thereby improving automated segmentation tools' applicability in clinical settings. The study's findings provide a roadmap for future research to adopt a data-centric approach in medical image AI model development.

## INTRODUCTION

Medical imaging combined with artificial intelligence-based approaches is playing a central role in diagnostic and therapeutic decision-making for numerous diseases [1]. Computed tomography (CT) scans, for instance, are commonly used for identifying abnormalities in the abdomen such as polycystic kidney disease (PKD) [2]. One crucial process in the examination of radiological images is segmentation, which entails demarcating regions of interest, such as organs or pathological features, for further analysis [3, 4, 5]. In recent years, automated segmentation, enabled by advancements in artificial intelligence (AI) and deep learning techniques, has been gaining increased attention as a tool to streamline the segmentation process and reduce the burden on radiologists [6].

Although AI-based segmentation models have demonstrated promising results in various applications, their performance often declines when applied to new or out-of-domain data [7]. This is particularly concerning in the medical imaging field, where models are expected to accommodate a wide range of imaging variations caused by differences in scanning protocols, patient populations, or disease states [8]. The ability of AI models to generalize from their training data to unseen data is thus critical for their practical applicability in the clinical setting.

This study assesses the generalization and out-of-domain performance of AI-based segmentation models trained on abdominal CT images, focusing on their adaptability to new image types and patient groups. We analyze the segmentation of kidneys, livers, and spleens in both non-contrast and contrast-enhanced scans, evaluating model performance via similarity metrics. We also explore the impact of dataset diversity (i.e. image acquisition and variations in patient phenotypes) on model generalization. Our goal is to enhance understanding of AI image segmentation's training strategies and inform future research for more robust, generalizable AI models.

## METHODS

### CT images characteristics

Our institution's review board approved this study, waiving the need for informed consent. We retrospectively collected 500 abdominal CT images from our radiology database. Scans were performed at our institution between January 2001 and December 2018. From this, four datasets of 100 images each were curated, including non-contrast and contrast-enhanced CT images of healthy controls and PKD patients. Additionally, two mixed datasets of 100 images were created: one for PKD with equal proportions of contrast and non-contrast images, and one with 25% random images from each homogeneous dataset. A separate dataset of 100 non-contrast CT images from PKD patients was reserved for testing.

**Reference standard**

The reference standard segmentations for comparing the model's segmentation performance was generated by a trained medical image analyst using the ITK-Snap software v3.8.0 [9] ([www.itksnap.org](www.itksnap.org)). ITK-Snap is a widely used software application that allows for precise manual segmentation of structures in 3D radiological images. Different labels were assigned to each organ using the annotation tool. Labels one and two were assigned to the right and left kidney, respectively, label three to the liver, and label four to the spleen.

**Segmentation model**

In this study, we utilized the nnU-Net model [10], an adaptive neural network architecture with state-of-the-art performance designed specifically for medical image segmentation. We used the '3D_fullres' implementation, which offers full resolution 3D processing of input data, and the 'nnUNetTrainerV2_noMirroring' trainer implementation, a specific variant of the trainer that excludes the use of mirror data augmentation.

We used the nnU-Net model to segment right and left kidneys, liver, and spleen, training it on six separate datasets. A five-fold cross-validation approach was used, with the final model being an ensemble of models trained on each fold. This ensemble model was tested on a separate dataset of 100 non-contrast PKD cases for an unbiased evaluation, and was compared to the recently released publicly available TotalSegmentator model [11] for performance benchmarking against a state of the art model.

**Statistical Analysis**

We evaluated model performance using Dice coefficient, Jaccard index, true positive rate (TPR), and precision to measure segmentation accuracy and overlap with the reference standard. Performance comparison between models trained on the different datasets and those on the in-domain data was done via a non-inferiority Wilcoxon signed-rank test. A preset non-inferiority margin was used for each metric, with a score within this margin deemed non-inferior. A one-sided 95% confidence interval was used for the difference in metrics, considering a p-value under 0.05 as statistically significant.

**RESULTS**

As detailed in **Table 1**, the various datasets' patient and image characteristics, including patient demographics, image resolution, and organ volume distributions, were analyzed. Examples from each of the datasets are visually represented in **Figure 1**. The mean age for the control group was 43 years, with a range of 18-73 years. For PKD patients, the mean age was 52 years, with a range of 18-91 years. The entire cohort had a mean age of 48 years. The majority of patients were female, accounting for 56.6% in the control group and 53% in the PKD group. There were slightly more males in the non-contrast PKD groups.

The quality of images was approximated by comparing in-plane resolution between the datasets and was similar across all datasets (approximately 0.76-0.79 mm. Slice thickness was smallest in the contrast control group (mean 2.56 mm) and largest in the non-contrast PKD group (mean 4.20 mm). Kidney and liver volumes were notably larger in the PKD groups than in the control groups due to cystic changes (note – spleen volumes were also slightly larger).

**Table 2** provides the summary statistics of similarity metrics for each organ, derived from models trained on different datasets and tested on a holdout test set of non-contrast PKD patient images. For all organs, models trained on the control datasets and the publicly available TotalSegmentator model performed inferiorly, while models trained on the PKD contrast, PKD mixed, and all mixed datasets performed non-inferiorly compared to the in-domain non-contrast PKD model.

The in-domain PKD model consistently performed the best across all organs, implying the importance of training a model on a dataset that accurately represents the target patient population. The control datasets, though precise, had low true positive rates, suggesting they struggled to accurately identify the organs of interest in the non-contrast PKD test set.
The publicly available TotalSegmentator model showed varying performance across different organs but performed poorly in organs with more diverse phenotypic variability, such as the right and left kidneys. This is likely due to the lack of training dataset diversity of the TotalSegmentator model.

**Figure 2** illustrates that the model trained on an equal split from each dataset was comparable in its overlap measure (Dice coefficient) to the model trained purely on in-domain data. Specifically, the diverse model achieved a Dice score of 0.96 ± 0.05 for the right kidney, while the in-domain model scored 0.97 ± 0.06, demonstrating non-inferiority of the diverse model. The Jaccard index, true positive rate, and precision all confirmed these findings. As anticipated, a drop in model performance occurred due to differences in organ appearance and was lowest when identifying PKD kidneys in non-contrast CT exams from the contrast control dataset model.

**DISCUSSION**

Models trained on diverse datasets show robustness across various clinical scenarios, potentially enhancing diagnostic accuracy and efficiency in medical imaging. To enhance precision in distinguishing left and right structures, specifically kidneys, we utilized the 'nnUNetTrainerV2_noMirroring' trainer, avoiding mirror augmentation due to human abdominal asymmetry and organ pair challenges.

However, our study has limitations. Our findings, based on specific abdominal CT images and centered on kidney, liver, and spleen segmentation, may not apply to other organs or imaging modalities. Similarly, our reliance on non-contrast and contrast-enhanced CT images may limit the generalizability to other imaging techniques like MRI or PET. The study's focus on PKD patients might bias the training datasets, affecting generalization to other patient populations or diseases. In addition, optimal dataset size is another important consideration described in prior work [12]. Furthermore, we did not analyze how different proportions of data types or sources impact model performance, leaving the optimal balance for model training unexplored.

Nonetheless, our results suggest that training medical image segmentation models on diverse data can improve generalization and out-of-domain performance, promoting a data-centric approach to developing medical image AI models for real-world effectiveness.

**Table Legends**

**Table 1.** Basic characteristics of the various datasets used in this study. Clinical characteristics include age and sex, while imaging characteristics denote in-plane resolution and slice thickness. Organ volumes, listed in cubic centimeters (cc), cover the right and left kidneys, total kidney, liver, and spleen. Each characteristic is reported as mean ± standard deviation, with the range in brackets.

**Table 2.** Similarity metric summary statistics for models trained with different datasets and tested on a hold test set comprising patients affected by polycystic kidney disease imaged with non-contrast CT (N=100). The evaluation metrics include the Dice coefficient, Jaccard index, True Positive Rate (TPR), and Precision. Each metric is provided for the segmentation of the right kidney, left kidney, liver, and spleen, with values expressed as mean ± standard deviation. These metrics provide insights into the accuracy and reliability of the segmentation results for different patient groups and tools.

**Figure Legends**

**Figure 1.** Example images from each of the datasets used to train the various segmentation models.

**Figure 2.** Box plot summary showing variation in test set performance between models trained with different patient phenotypes and imaging types. Metrics are shown for Dice, Jaccard, True Positive Rate (TPR), True Negative Rate (TNR), and Precision for the right kidney segmentation. Note similar performance for the 'PKD – NonCons' with the 'Full – Mix' dataset which contains only 25% in-domain data.

# REFERENCES

1. Thrall JH, Li X, Li Q, et al. (2018) Artificial Intelligence and Machine Learning in Radiology: Opportunities, Challenges, Pitfalls, and Criteria for Success. J Am Coll Radiol 15**:** 504-508. 10.1016/j.jacr.2017.12.026
2. Grantham JJ (2008) Clinical practice. Autosomal dominant polycystic kidney disease. N Engl J Med 359**:** 1477-85. 10.1056/NEJMcp0804458
3. Kline TL, Korfiatis P, Edwards ME, et al. (2017) Performance of an Artificial Multi-observer Deep Neural Network for Fully Automated Segmentation of Polycystic Kidneys. J Digit Imaging 30**:** 442-448. 10.1007/s10278-017-9978-1
4. Gregory AV, Anaam DA, Vercnocke AJ, et al. (2021) Semantic Instance Segmentation of Kidney Cysts in MR Images: A Fully Automated 3D Approach Developed Through Active Learning. J Digit Imaging 34**:** 773-787. 10.1007/s10278-021-00452-3
5. Korfiatis P, Denic A, Edwards ME, et al. (2022) Automated Segmentation of Kidney Cortex and Medulla in CT Images: A Multisite Evaluation Study. J Am Soc Nephrol 33**:** 420-430. 10.1681/ASN.2021030404
6. Chartrand G, Cheng PM, Vorontsov E, et al. (2017) Deep Learning: A Primer for Radiologists. Radiographics 37**:** 2113-2131. 10.1148/rg.2017170077
7. Kline TL (2019) Segmenting New Image Acquisitions Without Labels. 2019 IEEE 16th International Symposium on Biomedical Imaging (ISBI)**:** 330-333.
8. Oakden-Rayner L, Dunnmon J, Carneiro G,Re C (2020) Hidden Stratification Causes Clinically Meaningful Failures in Machine Learning for Medical Imaging. Proc ACM Conf Health Inference Learn (2020) 2020**:** 151-159. 10.1145/3368555.3384468
9. Yushkevich PA, Piven J, Hazlett HC, et al. (2006) User-guided 3D active contour segmentation of anatomical structures: Significantly improved efficiency and reliability. Neuroimage 31**:** 1116-1128. 10.1016/j.neuroimage.2006.01.015
10. Isensee F, Jaeger PF, Kohl SAA, Petersen J,Maier-Hein KH (2021) nnU-Net: a self-configuring method for deep learning-based biomedical image segmentation. Nat Methods 18**:** 203-+. 10.1038/s41592-020-01008-z
11. Wasserthal J, Breit H-C, Meyer MT, et al. (2022) TotalSegmentator: robust segmentation of 104 anatomical structures in CT images. Journal**:** arXiv:2208.05868. 10.48550/arXiv.2208.05868
12. Gottlich HC, Gregory AV, Sharma V, et al. (2023) Effect of Dataset Size and Medical Image Modality on Convolutional Neural Network Model Performance for Automated Segmentation: A CT and MR Renal Tumor Imaging Study. J Digit Imaging. 10.1007/s10278-023-00804-1

Table 1. Basic characteristics of the various datasets used in this study. Clinical characteristics include age and sex, while imaging characteristics denote in-plane resolution and slice thickness. Organ volumes, listed in cubic centimeters (cc), cover the right and left kidneys, total kidney, liver, and spleen. Each characteristic is reported as mean ± standard deviation, with the range in brackets.

| Properties | Controls - NonCons | Controls - Cons | PKD - NonCons | PKD - Cons | Test - PKD - NonCons |
|---|---|---|---|---|---|
| **Clinical Characteristics** | | | | | |
| Age (years) | 44 ± 13 [18 73] | 42 ± 13 [22 73] | 53 ± 15 [18 86] | 52 ± 17 [20 83] | 53 ± 15 [29 91] |
| Sex (M/F) | 41/59 | 46/54 | 59/41 | 29/71 | 53/47 |
| **Image Characteristics** | | | | | |
| In-Plane Resolution (mm) | 0.77 ± 0.08 [0.59 0.98] | 0.76 ± 0.09 [0.59 0.98] | 0.79 ± 0.09 [0.59 0.98] | 0.76 ± 0.09 [0.59 0.98] | 0.77 ± 0.08 [0.59 0.98] |
| Slice Thickness (mm) | 3.00 ± 1.03 [0.50 5.00] | 2.56 ± 0.91 [0.60 5.00] | 4.20 ± 1.19 [1.50 7.00] | 3.93 ± 1.46 [0.80 7.00] | 3.81 ± 1.30 [0.80 5.00] |
| **Organ Volumes** | | | | | |
| Right Kidney (cc) | 167 ± 34 [110 247] | 184 ± 37 [121 295] | 1140 ± 1273 [79 7538] | 495 ± 248 [121 1192] | 1103 ± 812 [111 3257] |
| Left Kidney (cc) | 171 ± 37 [107 269] | 184 ± 38 [108 293] | 1117 ± 1038 [108 4870] | 512 ± 293 [129 1545] | 1158 ± 834 [96 3367] |
| Total Kidney (cc) | 337 ± 69 [220 516] | 368 ± 73 [237 ± 588] | 2257 ± 2197 [188 9736] | 1007 ± 504 [257 2642] | 2262 ± 1605 [266 6078] |
| Liver (cc) | 1496 ± 341 [910 2516] | 1586 ± 335 [971 2638] | 2141 ± 1112 [864 7658] | 2398 ± 1794 [863 11884] | 2385 ± 1311 [952 7778] |
| Spleen (cc) | 212 ± 90 [73 528] | 239 ± 99 [67 ± 589] | 307 ± 212 [95 2018] | 304 ± 171 [35 1125] | 313 ± 157 [39 756] |

**Table 2.** Similarity metric summary statistics for models trained with different datasets and tested on a hold test set comprising patients affected by polycystic kidney disease imaged with non-contrast CT (N=100). The evaluation metrics include the Dice coefficient, Jaccard index, True Positive Rate (TPR), and Precision. Each metric is provided for the segmentation of the right kidney, left kidney, liver, and spleen, with values expressed as mean ± standard deviation. These metrics provide insights into the accuracy and reliability of the segmentation results for different patient groups and tools.

| Metric | Controls - NonCons | Controls - Cons | PKD - NonCons | PKD - Cons | PKD - Mix | All - Mix | TotalSegmentator |
|---|---|---|---|---|---|---|---|
| **Right Kidney** | | | | | | | |
| Dice | 0.64 ± 0.33 | 0.04 ± 0.16 | 0.97 ± 0.06 | 0.95 ± 0.06 | 0.97 ± 0.06 | 0.96 ± 0.05 | 0.45 ± 0.36 |
| Jaccard | 0.55 ± 0.36 | 0.03 ± 0.12 | 0.94 ± 0.08 | 0.92 ± 0.10 | 0.94 ± 0.08 | 0.94 ± 0.09 | 0.36 ± 0.33 |
| TPR | 0.56 ± 0.36 | 0.03 ± 0.13 | 0.98 ± 0.01 | 0.94 ± 0.08 | 0.98 ± 0.02 | 0.98 ± 0.02 | 0.37 ± 0.33 |
| Precision | 0.94 ± 0.16 | 0.60 ± 0.48 | 0.96 ± 0.08 | 0.97 ± 0.06 | 0.96 ± 0.08 | 0.96 ± 0.08 | 0.99 ± 0.10 |
| **Left Kidney** | | | | | | | |
| Dice | 0.62 ± 0.30 | 0.03 ± 0.12 | 0.97 ± 0.04 | 0.96 ± 0.06 | 0.97 ± 0.04 | 0.97 ± 0.07 | 0.45 ± 0.31 |
| Jaccard | 0.52 ± 0.33 | 0.02 ± 0.10 | 0.95 ± 0.06 | 0.93 ± 0.09 | 0.95 ± 0.06 | 0.94 ± 0.09 | 0.35 ± 0.29 |
| TPR | 0.53 ± 0.34 | 0.02 ± 0.10 | 0.98 ± 0.04 | 0.95 ± 0.09 | 0.98 ± 0.02 | 0.97 ± 0.07 | 0.35 ± 0.29 |
| Precision | 0.97 ± 0.10 | 0.54 ± 0.50 | 0.97 ± 0.05 | 0.98 ± 0.04 | 0.97 ± 0.06 | 0.97 ± 0.07 | 0.99 ± 0.01 |
| **Liver** | | | | | | | |
| Dice | 0.89 ± 0.11 | 0.79 ± 0.25 | 0.97 ± 0.02 | 0.96 ± 0.03 | 0.97 ± 0.01 | 0.97 ± 0.01 | 0.90 ± 0.09 |
| Jaccard | 0.82 ± 0.15 | 0.71 ± 0.29 | 0.94 ± 0.03 | 0.92 ± 0.05 | 0.94 ± 0.02 | 0.94 ± 0.02 | 0.83 ± 0.13 |
| TPR | 0.88 ± 0.14 | 0.73 ± 0.30 | 0.97 ± 0.03 | 0.96 ± 0.02 | 0.97 ± 0.02 | 0.97 ± 0.02 | 0.92 ± 0.13 |
| Precision | 0.92 ± 0.08 | 0.96 ± 0.05 | 0.97 ± 0.02 | 0.96 ± 0.05 | 0.97 ± 0.02 | 0.97 ± 0.02 | 0.90 ± 0.08 |
| **Spleen** | | | | | | | |
| Dice | 0.91 ± 0.12 | 0.90 ± 0.13 | 0.97 ± 0.04 | 0.95 ± 0.08 | 0.97 ± 0.03 | 0.96 ± 0.03 | 0.88 ± 0.13 |
| Jaccard | 0.85 ± 0.16 | 0.83 ± 0.17 | 0.93 ± 0.06 | 0.91 ± 0.11 | 0.93 ± 0.05 | 0.93 ± 0.06 | 0.80 ± 0.17 |
| TPR | 0.97 ± 0.02 | 0.89 ± 0.16 | 0.98 ± 0.02 | 0.98 ± 0.02 | 0.98 ± 0.02 | 0.98 ± 0.02 | 0.94 ± 0.05 |
| Precision | 0.87 ± 0.17 | 0.92 ± 0.10 | 0.96 ± 0.06 | 0.93 ± 0.11 | 0.96 ± 0.05 | 0.95 ± 0.05 | 0.84 ± 0.18 |

**Figure 1.** Example images from each of the datasets used to train the various segmentation models.

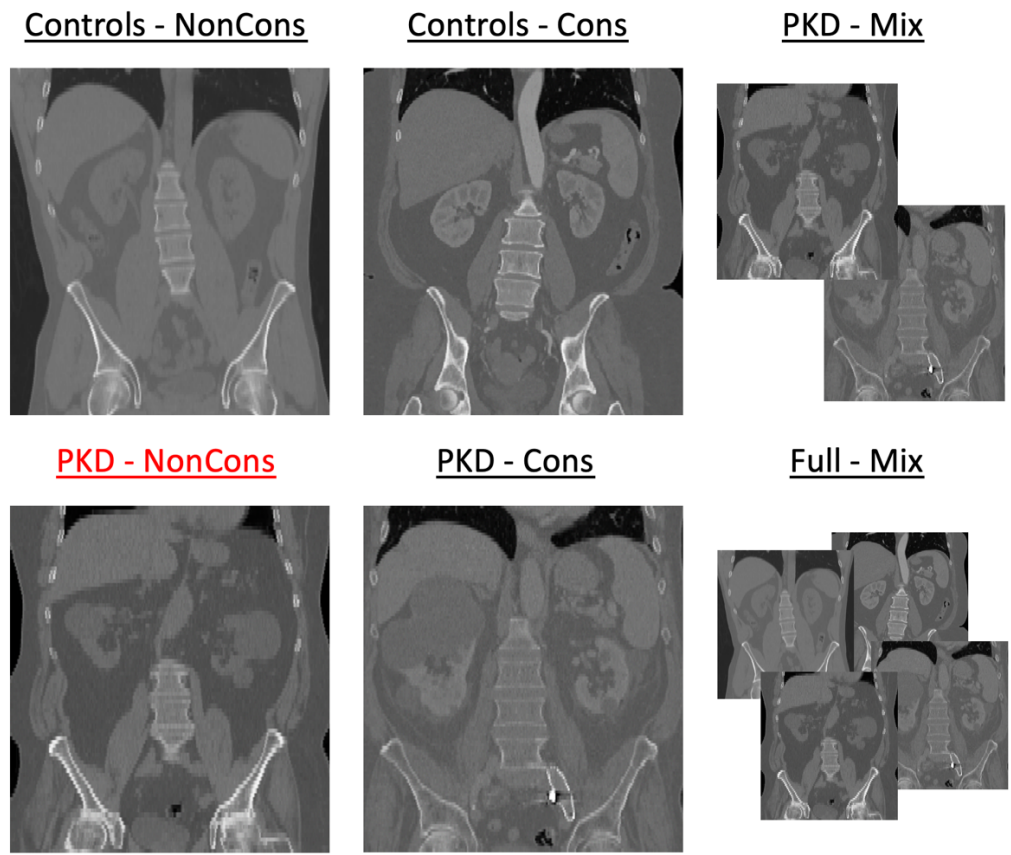

**Figure 2.** Box plot summary showing variation in test set performance between models trained with different patient phenotypes and imaging types. Metrics are shown for Dice, Jaccard, True Positive Rate (TPR), True Negative Rate (TNR), and Precision for the right kidney segmentation. Note similar performance for the 'PKD – NonCons' with the 'Full – Mix' dataset which contains only 25% in-domain data.

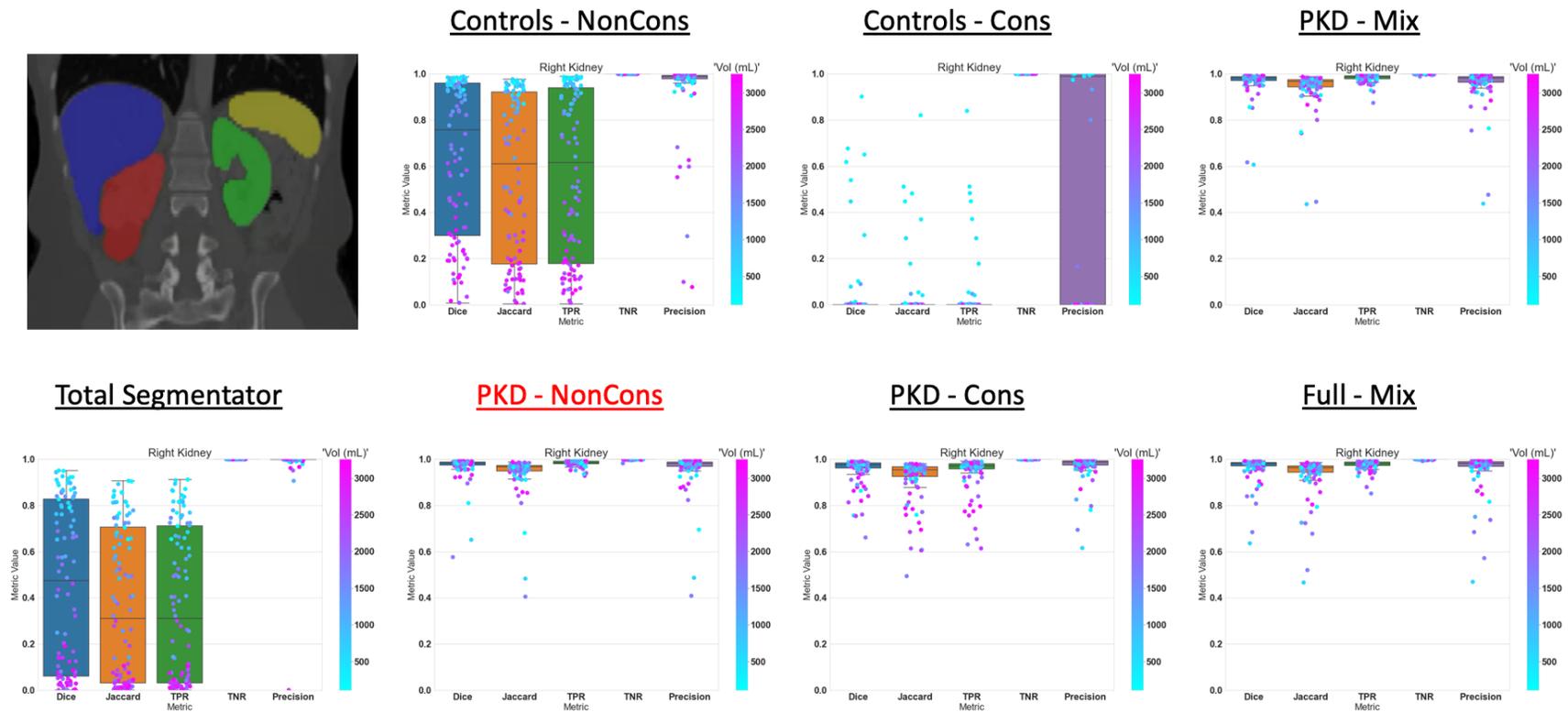